\documentclass[aps,superscriptaddress,showpacs,nofootinbib,twocolumn]{revtex4}

\begin{document}

\title{Convergent resummed linear $\delta$
expansion in the critical $O(N)$ $(\phi_i^2)^2_{3d}$ model}

\author{Jean-Lo\"{\i}c Kneur}
\affiliation{Laboratoire de Physique Math\'{e}matique et Th\'{e}orique - CNRS - UMR
5825 Universit\'{e} Montpellier II, Montpellier, France}

\author{Marcus B. Pinto}
\affiliation{Laboratoire de Physique Math\'{e}matique et Th\'{e}orique - CNRS - UMR
5825 Universit\'{e} Montpellier II, Montpellier, France}
\affiliation{Departamento de F\'{\i}sica,
Universidade Federal de Santa Catarina,
88040-900 Florian\'{o}polis, SC, Brazil}

\author{Rudnei O. Ramos}
\affiliation{Departamento de F\'{\i}sica Te\'orica,
Universidade do Estado do Rio de Janeiro,
20550-013 Rio de Janeiro, RJ, Brazil}

\begin{abstract}

The nonperturbative linear $\delta$ expansion (LDE) method is applied to
the critical $O(N)$ $\phi^4$ three-dimensional field theory which has
been widely used to study the critical temperature of condensation of
dilute weakly interacting homogeneous Bose gases. We study the higher
order convergence of the LDE as it is usually applied to this problem.
We show how to improve both, the large-$N$ and finite $N=2$, LDE results
with an efficient resummation technique which accelerates convergence.
In the large $N$ limit, it reproduces the known exact result within
numerical integration accuracy. In the finite $N=2$ case, our improved
results support the recent numerical Monte Carlo estimates for the
critical transition temperature of Bose-Einstein condensation.

\end{abstract}


\maketitle

Perturbative methods are well established tools to study diverse aspects
of physical systems. However, in many important problems one is not
allowed to use perturbation theory, or it just breaks down. In field
theory, a well known example concerns the description of physical
phenomena close to critical points of continuous or second order phase
transitions. In this situation the divergence of length and time scales
of the fluctuations, associated to infrared (IR) divergences and
critical slow-down, respectively, result in the singular behavior of
many physical quantities, like correlation lengths, susceptibilities and
many other parameters. In these cases one must recur to nonperturbative
methods, such as the renormalization group methods, the
$\epsilon$-expansion, the $1/N$ approximation and other techniques (for
a review, see Ref. \cite{zustin}).

A timely important problem associated to the breakdown of perturbation
theory near a critical point is the evaluation of the critical
transition temperature for a weakly interacting homogeneous Bose gas.
This apparently simple, but highly nontrivial task has been the source
of controversy for many years. Recently, its functional form was found
to be $T_c= T_0 \{ 1 + c_1 a n^{1/3} + [ c_2^{\prime} \ln(a n^{1/3})
+c_2^{\prime \prime} ] a^2 n^{2/3} + {\cal O} (a^3 n)\}$, where $T_0$ is
the ideal gas condensation temperature, $a$ is the $s$-wave scattering
length, $n$ is the density and $c_1, c_2^{\prime}, c_2^{\prime\prime}$
are numerical coefficients \cite {second}. However, a strong debate
concerns the values of the numerical coefficients, especially $c_1$,
which has been computed by different authors, using different methods.
{} Some analytical predictions included the self-consistent resummation
schemes ($c_1\simeq 2.90$) \cite {baymprl}, the $1/N$ expansion at
leading order ($c_1 \simeq 2.33$) \cite {baymN} and at next to leading
order ($c_1 \simeq 1.71$) \cite {arnold1} and also the linear
$\delta$-expansion (LDE) \cite {linear,odm} at second order ($c_1\simeq
3.06$) \cite {prb}. The numerical methods include essentially Monte
Carlo lattice simulations (MC). The most recent MC results are reported
by the authors of Ref. \cite{russos} ($c_1 = 1.29 \pm 0.05$) and of Ref.
\cite{arnold2} ($c_1 = 1.32 \pm 0.02$). The problem is that these
coefficients (except $c_2^\prime$) are sensitive to the IR physics at
the critical point and so, no perturbative approach can be used to
predict them. 

At first, one may believe that, due to the complex nature of the
problem, it is unlikely that a definitive analytical prediction of the
IR coefficients could be made. In general, higher order computations
quickly become prohibitively difficult with the traditional
nonperturbative analytical methods which, like the $1/N$ expansion, rely
on the resummation of an infinite number of particular contributions. At
the same time, the LDE great advantage is the fact that all
calculations, including renormalization, are performed in a perturbative
way. This means that one always deals with a finite set of contributions
at each order. This advantage is well illustrated in a recent work by
some of the present authors \cite{pra}, where the LDE was extended to
order-$\delta^4$. The results give strong indications that, when
properly applied, the LDE leads to very precise {\it analytical}
predictions for the nonperturbative coefficients $c_1$ and
$c_2^{\prime\prime}$. In fact, the results of Ref. \cite{pra} for these
coefficients, at second ($c_1\simeq 3.06, \; c^{\prime\prime}_2 \simeq
101.4$), third ($c_1 \simeq 2.45, \; c^{\prime \prime}_2 \simeq 98.2$)
and fourth orders ($c_1 \simeq 1.48, \; c^{\prime \prime}_2 \simeq
82.9$) seem to, roughly, converge to the precise MC results of Ref.
\cite{arnold2}, where $c^{\prime \prime}_2 = 75.7 \pm 0.4$ was also
predicted.

The important question regarding the LDE convergence properties has been
addressed in the context of the anharmonic oscillator (AO) at zero
temperature where rigorous proofs have been produced \cite {ian}. Those
proofs have been extended to the finite temperature domain by Duncan and
Jones \cite {dujo} who considered the AO partition function. Very
recently, the convergence proof has been extended to renormalizable
quantum field theories at zero temperature \cite {jldamien}. However, it
would be interesting to probe convergence in the vicinity of a phase
transition such as for the Bose-Einstein condensation (BEC) model 
considered here. In most
applications one can establish simple relations in between the LDE and
other nonperturbative methods already at order-$\delta$ where one-loop
diagrams are present. In fact, one can show that in those cases the LDE
either exactly reproduces $1/N$ results or produce very close numerical
estimates \cite {ian,sunil}. Here, the BEC problem poses an additional
difficulty since the first non-trivial contributions start at the
two-loop level in the self-energies. This is a consequence of the
Hugenholtz-Pines theorem which eliminates the one-loop momentum
independent contributions. In this case, it is not easy to establish
simple analytical relations like those given in Refs. \cite {ian,sunil}
and the problem must be treated differently. 

Braaten and Radescu in Ref. \cite
{braaten} have recently re-visited the LDE application to the BEC
problem also considering the convergence problem. One of the differences
between our present work and also Refs. \cite {prb,pra} from their
approach regards the choice of physical quantity to be extremized.
Moreover, the coefficient $c_2^{\prime \prime}$ and convergence
acceleration, considered by us, were not investigated in Ref. \cite
{braaten}. We discuss both approaches, in more details, in a longer
companion paper \cite{new}.

In this Letter, our aim is to mainly investigate the convergence of the
LDE, in the critical $O(N)$ $(\phi_i^2)^2_{3d}$ model, by considering
its behavior at the large-$N$ as well as at the finite $N=2$ limits.
This study will be combined with a simple, but powerful, all order LDE
resummation technique \cite{gn2} so that we can explicitly show
convergence.

At the critical point, one can describe a weakly interacting dilute
homogeneous Bose gas by an effective action analogous to a 
$O(2)$ scalar field model in three-dimensions given by

\begin{equation}
S_{\phi}=  \int d^3x \left ( \frac {1}{2} ( \nabla \phi )^2 +
\frac {1}{2} r
\phi^2 + \frac {u}{4!} \phi^4
\right ) \;,
\label{action2}
\end{equation}
where $\phi$ is a two-component real scalar field.
The parameters $r$ and $u$ are related to the original
parameters of the nonrelativistic action, the chemical
potential $\mu$, $a$, atomic mass $m$ and temperature $T$ by
$r=-2m\mu$ and $u=48 \pi a mT$ \cite{baymN,arnold1}.
The leading order coefficient of the critical temperature
shift can be expressed as \cite{baymprl} 
$c_1 = - 128 \pi^3
[\zeta(3/2)]^{-4/3} {\Delta \langle \phi^2 \rangle}/u$,
where $\Delta
\langle \phi^2 \rangle = \langle \phi^2 \rangle_u - \langle \phi^2
\rangle_0$.
The subscripts $u$ and $0$ mean that the field fluctuations
are to be evaluated in the presence and in the absence of interactions
respectively. 

The implementation of LDE within this model is reviewed in  previous
applications \cite{prb,pra} (see also Ref. \cite{prd}).
In practice, one considers the  original theory Eq. (\ref{action2}) adding
a quadratic (in the fields) term $1/2(1-\delta)\eta^2
\phi^2$ where $\eta$ is an arbitrary mass parameter. At the same time $u
\to \delta u$ and $r \to \delta r$. One ends up
with an interpolated theory described by the propagator
$(q^2+\eta^2)^{-1}$ and vertices $-\delta u$ , $\delta \eta^2$ and
$-\delta r$. A physical quantity $\Phi^{(k)}$ is then evaluated to an
order-$\delta^k$ using these new {}Feynman rules and following the
program, including renormalization, of ordinary perturbation theory.
This method operates for any $N$ and is free from IR divergences.
Nonperturbative results for the original ($\eta$-independent) theory are
obtained by optimizing $\Phi^{(k)}$ with respect to $\eta$ at $\delta=1$. One
possible optimization procedure, adopted in a part of 
this work, is the Principle of Minimal Sensitivity (PMS)
which requires $ d \Phi^{(k)} /d \eta =0$ \cite {pms}.
{}For comparison and to check the consistency of the
optimization procedure for reproducing physically sensible results,
we also consider the alternative optimization scheme
known as the {}Fastest Apparent Convergence criterion (FAC)
where one requires  that, at a given order $k$,
the $k$-th coefficient of the perturbative expansion
be zero. As we will see, both methods lead to equivalent
results.

Let us first consider, in the large-$N$ limit, the usual analytical way
\cite {prb,pra} to define a non-trivial LDE, by performing an order by
order perturbative evaluation of $\langle \phi^2 \rangle_u^{(k)}$ which,
moreover, has the advantage of being straightforwardly generalized for
arbitrary $N$ values. The convergence property issues can be studied by
comparing the results with the ``exact" $1/N$ result \cite{baymN}
$c_1=2.328$. The details of such a calculation follow from
the methods used in Ref. \cite{pra} and are given in the accompanying
paper \cite{new}. One obtains

\begin{equation}
\langle \phi^2 \rangle_u^{(20)} \!=\! 
- \frac {N \eta^*}{4\pi} + \delta \frac {N u}{3}
\sum_{i=1}^{19} C_i \left ( - \frac {\delta u N}{6 \eta^*} \right )^i  
+ {\cal O}(\delta^{21}),
\label{exp20}
\end{equation}
where the coefficients are given by $C_1=7.249\times10^{-5}$, $C_2=2.032
\times 10^{-6}$, $C_3=6.400\times 10^{-8}$, $C_4=2.080 \times 10^{-9}$,
$C_5=5.021 \times 10^{-11}$, $C_6 = 2.760 \times 10^{-12}$, $C_7 =3.580
\times 10^{-14}$, $C_8=6.500 \times 10^{-16}$, $C_9= 1.090 \times
10^{-17}$, $C_{10}= 1.040 \times 10^{-19}$, $C_{11}= 7.030 \times
10^{-22}$, $C_{12}= 2.810 \times 10^{-24}$, $C_{13}= 7.300 \times
10^{-26}$, $C_{14}= 2.800 \times 10^{-28}$, $C_{15}= 6.000 \times
10^{-31}$, $C_{16}= 5.780 \times 10^{-33}$, $C_{17}= 1.100 \times
10^{-35}$, $C_{18}= 1.130 \times 10^{-37}$ and $C_{19}= 1.390 \times
10^{-40}$. All coefficients for $i \ge 2$ in Eq.~(\ref{exp20}) were
obtained from $i$-dimensional integrals over {}Feynman parameters, that
we have performed by using the well-known Monte Carlo multidimensional
integration routine VEGAS \cite{vegas}. We should note that for such
high dimension and complicated integrals, the Monte-Carlo statistical
integration error cannot be reduced well below the percent level and
this will be reflected on our final results.

Here, as in the original LDE applications to the BEC problem
\cite{prb,pra}, our strategy is to evaluate $\langle \phi^2
\rangle_u^{(k)}$ perturbatively to order-$\delta^k$ and then to
extremize this quantity following the PMS. By setting $u=0$ in the optimal
$\langle \phi^2 \rangle_u^{(k)}$ one immediately obtains the optimal
$\langle \phi^2 \rangle_0^{(k)}$ and, at the same time, the optimal
$\Delta \langle \phi^2 \rangle^{(k)}$. By applying the PMS to this
quantity, at each order, one obtains solutions that can be grouped
into complex families whose first member is real as discussed in the 
anharmonic oscillator analogous
studies (see Bellet, Garcia and Neveu in Refs. \cite {ian}) 
and in Ref. \cite {pra}. To
order-$\delta^{20}$, our results are shown in Table I, together with
those obtained by the FAC optimization procedure, for 
the best converging family of solutions (see Ref. \cite{new}
for details). In the very last line of
Table I, we indicate the corresponding statistical integration accuracy
as provided by VEGAS. 
Note that both, the PMS and FAC, results converge to similar values,
attesting that our results are not merely an
artifact of the optimization procedure.

\begin{table}
\begin{center}
\caption{PMS, FAC and CIRT-PMS results for $c_1$ at large-$N$, 
at different orders $k$, obtained
with the best converging families (real part) toward the exact result
$c_1=2.328$.}
\begin{tabular}{c||c|c|c}
\hline
$ k $ &  PMS  & FAC & CIRT-PMS \\
\hline\hline
2   &  4.326             & 4.996 & -- \\
\hline
3  & $3.760 $ & $3.774 $ & --\\
\hline
4 & $4.400 $ & $3.298 $ & $4.886$ \\
\hline
5 & $3.260 $ & $3.126 $ & $4.258 $  \\
\hline
6 & $3.000 $& $2.940 $ & $3.622 $ \\
\hline
7 & $2.852$ & $2.810 $ & $2.450 $ \\
\hline
8 & $2.760 $ & $2.738 $ &$ 2.310 $ \\
\hline
9 & $2.700 $ & $2.684$ & $2.340 $  \\
\hline
10 & $2.652$& $2.638 $ &$2.346 $ \\
\hline
11 & $2.600 $& $2.604 $ &$2.346 $  \\
\hline
12 & $2.580 $& $2.574 $ &$2.346 $  \\
\hline
13 & $2.540 $&$2.552 $ &$2.346 $ \\
\hline
14 & $2.538  $& $2.532$ & $2.346 $ \\
\hline
15 & $2.520 $& $2.516 $& $2.346 $ \\
\hline
16 & $2.504 $& $2.504 $& $2.346 $  \\
\hline
17 & $2.488$& $2.492 $&$2.346 $  \\
\hline
18 & $2.480 $&$2.482 $&$2.346 $ \\
\hline
19 & $2.476 $& $2.474 $&$2.346 $  \\
\hline
20 & $2.468 \pm 0.025$& $2.466 \pm 0.025$ & $2.346 \pm 0.023$  \\
\hline
\end{tabular}
\end{center}
\end{table}

As far as the predictions of Ref. \cite {braaten} are concerned we note
that our results, optimized via PMS or FAC, are also very stable. Like
those authors, at about $18^{\rm th}$ order we achieve $\sim 7 \%$
accuracy. Let us now present a way of improving the above results with
an efficient LDE resummation technique. Performing the usual LDE
interpolation with $\eta^* = \eta (1 - \delta)^{1/2} $ and $u \to \delta
u$, expanded to order $p$, defines a partial sum $\Phi^{(p)}(\eta,u,\delta)
\equiv \sum^p_{n=0} s_n \delta^n$, which for $\delta\to 1$ is given
formally, from the simple pole residues, as

\begin{equation}
\Phi^{(p)}(\eta,u,\delta\to 1) =\frac{1}{2\pi\,i}\, \oint
d\delta\,\frac{\delta^{-p-1}}{1-\delta}\,\Phi(\eta,u,\delta)\;, 
\label{cont1}
\end{equation}
where the anticlockwise contour encircles the origin. Now, one performs
a change of variables \cite{gn2} for the relevant $\delta\to 1$ 
limit: $\delta \equiv 1-v/p$, together with a similarly
order-dependent rescaling of the arbitrary mass parameter, $\eta \to
\eta \;p^{1/2}$, where the power $1/2$ is dictated by the scalar mass
interaction term. {}For $p\to\infty$ this resummation can be summarized as
the replacement $\eta^* \to \eta v^{1/2}$, followed by the contour
integration 

\begin{equation}
\langle \phi^2 \rangle_{\stackrel{p\to\infty}{\delta \to 1}} =
\frac{1}{2\pi i}\, \oint \frac{dv}{v}\: \exp(v) \:
\langle  \phi^2( \eta^* \to \eta  v^{1/2}) \rangle,
\end{equation}
where the ``weight" $\exp(v)/v$ originates from
$d\delta \,(1-\delta)^{-1} \to dv/v$; 
$\lim_{\,p\to\infty}(1-v/p)^{-p-1}=\exp(v) $, and
the original contour was deformed to encircle the branch cut $Re[v]<0$. 
Noting that the LDE produces a  power series in $u^{(k+1)}/\eta^{k}$ and
by using

\begin{equation}
\oint dv \exp(v) \: v^a = 2\pi\,i /\Gamma(-a)\,,
\end{equation}
one sees that the main effect of this resummation is to divide the
original expansion coefficients at order $k$ by terms $\Gamma(1+k/2)
\sim (k/2)!$ for large $k$. This damping of the perturbative
coefficients at large order, as implied by this specific resummation, is
rather generic and was exploited recently in the completely different
context of asymptotically free models \cite{jldamien} where it was shown
to accelerate convergence of the LDE. When applied to the anharmonic
oscillator, it is in fact (asymptotically) equivalent to the more direct
LDE resummation with an order-dependent rescaling of the arbitrary mass,
as employed in some of the Refs.~\cite{ian} to establish rigorous
convergence of the LDE for the oscillator energy levels, which is itself
an extension of the order-dependent mapping (ODM) resummation technique
\cite{odm}. The contour integral resummation is however very convenient
since, algebraically, it is simpler than the direct LDE summation, in
particular to recover the original theory for $\eta^*\to 0$. The results
obtained through this contour integral resummation technique (CIRT)
applied to Eq. (\ref{exp20}) are also shown in Table I. {}Fast
convergence happens already at order-$\delta^{10}$, within $\sim 1 \%$
of the exact large-$N$ result for $c_1$. Note the complete stability of
results from this order onwards. Within the intrinsic numerical
integration error involved in the computation of the coefficients in Eq.
(\ref{exp20}) we can conclude on the actual convergence of the CIRT
towards the exact $1/N$ result. Note how the CIRT-PMS predictions
quickly become more accurate, than the ordinary PMS (FAC) results, when
more approximants are considered.

{}We finally turn to the finite $N=2$
case, for which the quantity $\langle \phi^2 \rangle_u^{(k)}$ has been
evaluated, up to order-$\delta^4$, in Ref. \cite {pra}. Its perturbative
expansion is

\begin{equation}
\langle \phi^2 \rangle_u^{(4)}= - \frac { \eta^*}{2\pi} + \delta u
\sum_{i=1}^{3} K_i \left ( - \frac {\delta u}{\eta^*} \right )^i +
{\cal O}(\delta^5)  \;,
\label{expNF}
\end{equation}
where the coefficients are given by $K_1=3.222\times10^{-5}$, $K_2=1.524
\times 10^{-6}$ and $K_3=1.042\times 10^{-7}$. The contour integral
technique applied at large-$N$ also improves the finite $N=2$ result of
Ref. \cite{pra} showing that the complex family of solutions has real
parts that converge to $c_1 = 1.15 \pm 0.01$, whereas the most recent
Monte Carlo predictions are \cite{russos} $c_1 = 1.29 \pm 0.05$ and 
\cite{arnold2} $c_1 = 1.32 \pm
0.02$. Apart from $\langle \phi^2 \rangle^{(k)}$, the quantity $\delta
r_c^{(k)}= - \Sigma_{\rm ren}^{(k)}(0)$ also enters the evaluation of
the order-$a^2$ coefficient $c_2^{\prime \prime}$ which appears in the
$T_c$ expansion \cite {second}. For $N=2$ its order-$\delta^4$
perturbative evaluation is \cite {pra}

\begin{eqnarray}
\delta r_c^{(4)}&=&  \delta \frac {u \eta^*}{6\pi}
+ \delta^2 u^2 A_2 \left [ \ln
\left ( \frac{M}{\eta^*} \right ) - 0.59775 \right ]  \nonumber \\
&-&\delta^3 \frac {u^3}{\eta^*} A_3 +
\delta^4 \frac {u^4}{(\eta^*)^2} A_4  +  {\cal O} (\delta^5) \;\;,
\label{rcNF}
\end{eqnarray}
where the coefficients are $A_2=1.407 \times 10^{-3}$, $A_3=8.509 \times
10^{-5}$ and $A_4=3.523\times 10^{-6}$. Treating Eq. (\ref {rcNF}) with
the CIRT one obtains the result ${\rm Re}\;[ r_c^{(4)}(M=u/3)]=0.0010034
u^2$ which, together with the CIRT improved $\langle \phi^2
\rangle_u^{(4)}$ result, leads to $c_2^{\prime \prime}= 84.9 \pm 0.8$
\cite {new}, while the Monte Carlo estimate is $c_2^{\prime \prime}=
75.7 \pm 0.4$ \cite {second} . Note that the scale $M=u/3$ was
originally chosen in those Monte Carlo applications.

In conclusion, the LDE has been applied successfully to many different
problems in field theory where standard perturbation theory does not
apply. But despite of its successes, its applicability to higher orders
and the study of its convergence properties in field theory, beyond the
anharmonic oscillator problem \cite{ian}, have proven to be a challenge.
Here, we have studied the convergence of the LDE as applied to the
critical point of Bose-Einstein condensation. We have shown that the
perturbative LDE for the large-$N$ case converges towards the exact
result, once resummed to all orders. We have also used the LDE to
explicitly evaluate, in the large-$N$ limit, the coefficient $c_1$ to
order-$\delta^{20}$ using two different optimization procedures (PMS and
FAC), leading to equivalent converging results, but in both cases $\sim
7 \%$ accuracy is only achieved at about order-$\delta^{18}$. Then, we
have shown how the powerful contour integral resummation technique
(CIRT) accelerates convergence already at order-$\delta^{10}$ within $1
\%$ accuracy or less, which appears in fact only limited by the
intrinsic accuracy of the Monte-Carlo integration used to evaluate the
coefficients of the relevant series. This same technique was extended to
the relevant finite $N=2$ case where the recent order-$\delta^4$ results
for $c_1$ and $c_2^{\prime \prime}$ \cite {pra} were improved. It is
worth remarking that, the $c_1$ values $3.06$, $2.45$ and $1.48$
obtained at orders-$\delta ^2$, $\delta^3$ and $\delta^4$ in Ref. \cite
{pra} are close to $2.90$, $2.33$ and $1.71$ obtained by resumming
``setting sun" contributions \cite {baymprl}, leading order \cite
{baymN} and next to leading order $1/N$ \cite {arnold1} contributions,
respectively. The similarity between the LDE values at a given order and
the values produced by each one of those analytical nonperturbative
approximations should come as no surprise if one considers the type of
graphs resummed by the LDE, at each order. At order-$\delta^2$, only
``setting sun" contributions are considered while more typical $1/N$
leading order contributions appear at order-$\delta^3$. At
order-$\delta^4$, graphs which would belong to the $1/N$ at next to
leading order also contribute. This simple consideration shows that the
hierarchy of LDE numerical values is not a mere coincidence but a
consequence of the type of graphs considered at each order. The
numerical differences are due to the fact that the LDE actually mixes up
those contributions, at each order in $\delta$, irrespective of their
$1/N$ order. Here, apart from supporting the recent Monte Carlo
predictions, our improved results reinforce the potential of the LDE as
a powerful tool to treat nonperturbative problems in field theory.

M.B.P. and R.O.R. were partially supported by Conselho
Nacional de Desenvolvimento Cient\'{\i}fico e Tecnol\'ogico
(CNPq-Brazil).


\begin{thebibliography}{99}

\bibitem{zustin}J. Zinn-Justin, {\it Quantum Field Theory and Critical 
Phenomena} (Oxford University Press, 1996).

\bibitem{second} P. Arnold, G. Moore and B. Tom\'{a}sik, 
Phys. Rev. {\bf A65}, 013606 (2002).

\bibitem{baymprl} G. Baym, {\it et. al}, Phys. Rev. Lett. {\bf 83}, 1703 (1999).

\bibitem{baymN} G. Baym, J.-P. Blaizot and J. Zinn-Justin, Europhys.
Lett. {\bf 49}, 150 (2000).

\bibitem{arnold1} P. Arnold and B. Tom\'{a}sik, Phys. Rev. {\bf A62},
063604 (2000).

\bibitem{linear} A. Okopi\'{n}ska, Phys. Rev. {\bf D35}, 1835 (1987);
A. Duncan and M. Moshe, Phys. Lett. {\bf B215}, 352 (1988);
V.I. Yukalov, Mosc. Univ. Phys. Bull. {\bf 31}, 10 (1976);
Teor. Mat. Fiz. {\bf 28}, 92 (1976).

\bibitem{odm}R. Seznec and J. Zinn-Justin, J. Math. Phys. {\bf 20} 1398 (1979).

\bibitem{prb}F. F. de Souza Cruz, M. B. Pinto and R. O. Ramos,
Phys. Rev. {\bf B64}, 014515 (2001); Laser Phys. {\bf 12}, 203 (2002).

\bibitem{russos} V. A.  Kashurnikov, N. V. Prokof'ev and B. V. Svistunov,
Phys. Rev. Lett. {\bf 87}, 120402 (2001).

\bibitem{arnold2} P. Arnold and G. Moore, Phys. Rev. Lett. {\bf 87},
120401 (2001); Phys. Rev. {\bf E64}, 066113 (2001).

\bibitem{pra}F. F. de Souza Cruz, M. B. Pinto, R. O. Ramos and P. Sena,
Phys. Rev. {\bf A65}, 053613 (2002).

\bibitem{ian}C. M. Bender, A. Duncan and H. F. Jones, Phys. Rev. {\bf D49},
4219 (1994);
H. Kleinert and W. Janke, Phys. Lett. {\bf A206}, 283 (1995);
R. Guida, K. Konishi and H. Suzuki, Ann. Phys. (NY) {\bf 241},
152 (1995); {\it ibid.} {\bf 249},
109 (1996); B. Bellet, P. Garcia and A. Neveu,
Int. J. of Mod. Phys. {\bf A11},
5587 (1997); {\it ibid.} {\bf A11}, 5607 (1997).

\bibitem{dujo}A. Duncan and H.F. Jones, Phys. Rev.
{\bf D47}, 2560 (1993).

\bibitem{jldamien} J.-L. Kneur and D. Reynaud, 
Eur. J. of Phys. {\bf C} (2002), in press (hep-th/0107073);
Phys. Rev. {\bf D} (2002), in press (hep-th/0205133).

\bibitem{sunil}S. K. Gandhi, H. F. Jones and M. B. Pinto, 
Nucl. Phys. {\bf B359}, 429 (1991).

\bibitem{braaten} E. Braaten and E. Radescu, cond-mat/0206186; hep-ph/0206108.

\bibitem{new}J.-L. Kneur, M. B. Pinto and R. O. Ramos,  cond-mat/0207295.

\bibitem{gn2} C.~Arvanitis, F.~Geniet, M. Iacomi, J.-L.~Kneur and A.~Neveu,
Int. J. Mod. Phys. {\bf A12}, 3307 (1997).

\bibitem{prd}M. B. Pinto and R. O. Ramos, Phys. Rev. {\bf D60 },
105005 (1999); {\it ibid.} {\bf D61} 125016 (2000).

\bibitem{pms}P. M. Stevenson, Phys. Rev. {\bf D23}, 2916 (1981).

\bibitem{vegas}  G. P.~Lepage, J.~Comp.~Phys. {\bf 27}, 192 (1978).


\end{thebibliography}
\end{document}